# Astro2020 Science White Paper

# Science opportunities enabled by the era of Visible Band Stellar Imaging with sub-100 μarc-sec angular resolution.

**Thematic Areas:** ☒ Planetary Systems   ☒ Star and Planet Formation

☒ Formation and Evolution of Compact Objects   ☐ Cosmology and Fundamental Physics

☒ Stars and Stellar Evolution   ☒ Resolved Stellar Populations and their Environments

☐ Galaxy Evolution   ☐ Multi-Messenger Astronomy and Astrophysics

**Principal Author:**
Name:  D. Kieda
Institution: University of Utah
Email: dave.kieda@utah.edu
Phone: 801-581-6926

**Co-authors:** (names and institutions)
Monica Acosta, *Instituto de Astrofísica de Canarias*
Anastasia Barbano, *DNPC, Universite de Geneve*
Colin Carlile, *Lund University*
Michael Daniel, *Center for Astrophysics | Harvard & Smithsonian*
Dainis Dravins, *Lund University*
Jamie Holder, *University of Delaware*
Nolan Matthews, *University of Utah*
Teresa Montaruli, *DNPC, Universite de Geneve*
Roland Walter, *DNPC, Universite de Geneve*
Luca Zampieri, *INAF-Osservatorio Astronomico di Padova*

**Abstract**:
This white paper briefly summarizes stellar science opportunities enabled by ultra-high resolution (sub-100 μ arc-sec) astronomical imaging in the visible (U/V) wavebands. Next generation arrays of Imaging Cherenkov telescopes, to be constructed in the next decade, can provide unprecedented visible band imaging of several thousand bright (m< 6), hot (O/B/A) stars using a modern implementation of Stellar Intensity Interferometry (SII). This white paper describes the astrophysics/astronomy science opportunities that may be uncovered in this new observation space during the next decade.

## Introduction

Much of the progress in astronomy is driven by improved imaging. Stars visible to the naked eye typically have angular diameters on the order of one milliarcsecond (mas), while a handful of red supergiants extend for a few tens of mas. If we take the Sun as our guide the scale of granular convection is of order $10^{-3}$ the solar radius [1], with supergranulation features a factor ~30 larger providing information on the magnetic field structure in the photosphere [2]. Revealing details across and outside stellar surfaces therefore requires imaging with resolution measured in microarcseconds (µas). For optical imaging, the Rayleigh criterion implies a requirement of kilometer-scale optical interferometers to enable meaningful surface imaging of most main sequence stars. Astronomical imaging with multi-kilometer baseline interferometry has previously been achieved only by radio interferometers operating between Earth and deep space, and has not yet been feasible for optical astronomy, but will be in the coming decade (Figure 1).

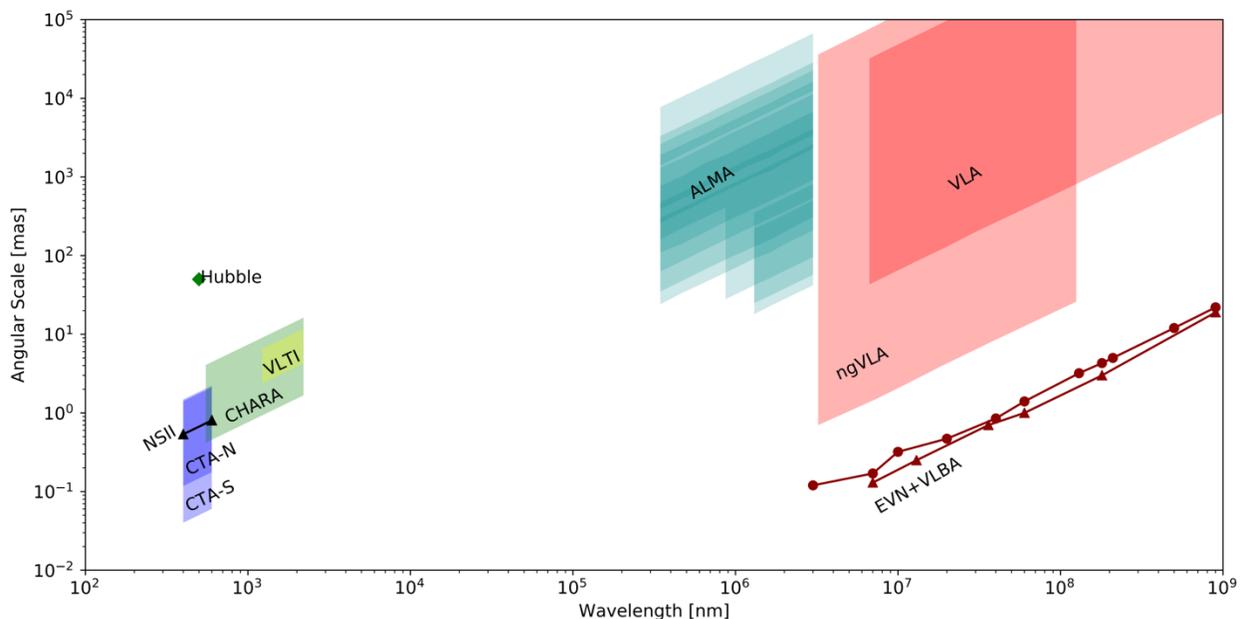

*Figure 1:* Angular resolution as a function of wavelength for radio arrays, mm/sub-mm arrays, and optical interferometer arrays (VLTI, CHARA). The angular resolution of the Hubble telescope is also provided for reference. NSII refers to the Narrabri Stellar Intensity Interferometer [3]. CTA-N and CTA-S refer to the resolution expected for a baseline SII implementation using a typical layout of the future CTA Northern Site Array and CTA Southern Site Array [4, 5], respectively.

SII exploits a second-order effect of light waves by measuring the quantum correlations in light intensity fluctuations from a source that is being observed by two or more separated telescopes. The correlation in light intensity fluctuations, measured as a function of telescope separation, measures the interferometric squared visibility, which is related to the angular brightness distribution of the source. SII is practically insensitive to either atmospheric turbulence or telescopic optical imperfections, enabling practical observations at very long (kilometer scale) baselines at short optical (U/V) wavelengths. This combination enables angular resolutions below 100 µas. SII telescopes only require interconnection using standard commercial fiber optic cable, allowing efficient construction of an array of one hundred or more optical telescopes distributed over many baseline scales, thereby fully sampling the Fourier image plane of the stellar source.

To appreciate the true meaning of such optical resolution, Figure 2 shows an 'understandable' type of object: solar-type phenomena projected onto the disk of a nearby star. A hypothetical exoplanet is shown in transit across the star Sirius. The planet's size and oblateness were made equal to that of Jupiter, including its four Galilean moons, and fitted with a Saturn-like ring. The stellar surface is assumed to be surrounded by a solar-type chromosphere, shining in a reddish emission-line, with protruding prominences and eruptions. This figure is not a simulation of any specific observation but an image matched to that angular resolution which is expected to be reached by using a significant subset of the telescopes planned for a future kilometer-scale array of ~10 meter diameter class optical telescopes, such as the Cherenkov Telescope Array (CTA) [4,5]. A discussion of the technical status and capabilities of such arrays for SII imaging will be presented in detail in a forthcoming reference document (to be submitted as an Astro 2020 APC white paper).

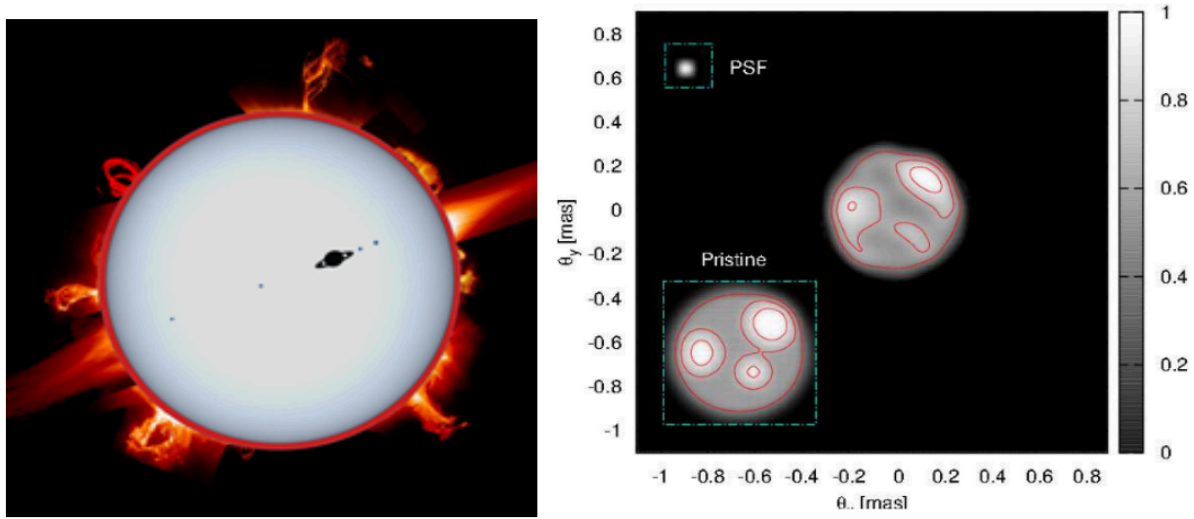

*Figure 2:* Illustrations of sub-100 µas optical resolution Left: A representative image of a transit of a hypothetical exoplanet across the disk of the relatively nearby star Sirius (6 mas diameter) imaged at 40 µas resolution. An active solar-type chromosphere is assumed above the stellar surface [6]. Right: A simulation of a $m_v$ = 3 star at 6000 K with hotspots between 6500K-6800K for a 10-hour observation and a simple CTA array layout. Reproduced from [7].

**Unexplored Discovery Space** Figure 3 plots all currently available directly measured stellar angular sizes [8] as a function of distance. The measurements of 32 stellar angular diameters by the Narrabri Stellar Intensity Interferometer (NSII) [3], with baselines up to ~180m, was the leading catalogue of angular size measurements ≤1 mas for many years, with the catalog more recently expanded by phase/amplitude interferometer observations[9]. The smallest angular scales have recently been explored through serendipitous asteroid occultation diffraction patterns observed with the VERITAS array of Imaging Air Cherenkov Telescopes (IACTs). It is clear that imaging stellar surfaces of 1-100 solar radius stars requires telescope baselines greater than several hundred meters. In the next decade kilometer-baseline IACT/intensity interferometry arrays, will image numerous stars in an unexplored parameter space, resulting in a comprehensive catalog of resolved stars over a broad range of distances, extending down to solar radius size.

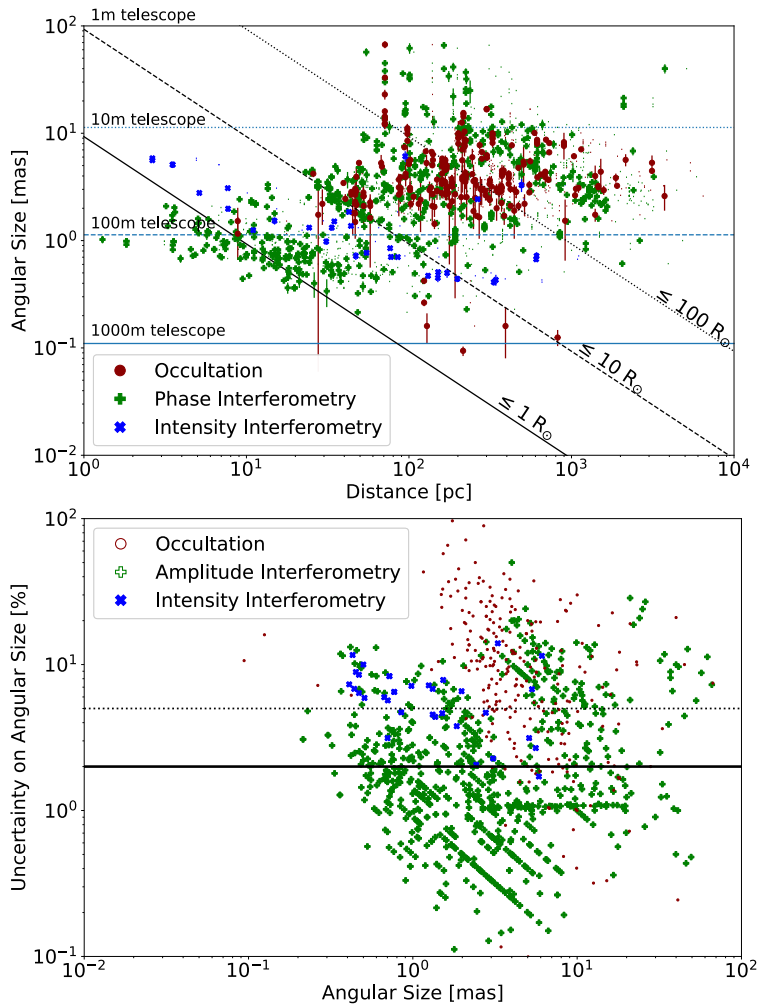

*Figure 3:* Top: Distribution of measured stellar angular sizes as a function of distance for existing measurements using asteroid occultation, phase interferometry, and intensity interferometry. Bottom: Uncertainty in angular size as a function of angular size for the three observation techniques.

**Limb Darkening** Limb darkening reduces the brightness of the edge of the stellar image compared to the central part of the disk due to the change in the optical depth of the photon path. Limb darkening is important for stellar modelling, and it also affects our understanding of transiting exoplanet lightcurves. Limb darkening is a limiting uncertainty in determining the exoplanet radius [10] and is also a potential systematic bias in determining the properties of exoplanet atmospheres [11]. High precision SII limb darkening measurements (< 5%) on suitable stars with exoplanets (e.g. HD209458-like) appear to be feasible using km-baseline observations. The resulting measurements can eliminate the use of model dependent limb darkening free parameters in stellar atmospheric fits [12].

**Starspots** Flaring in main sequence stars is believed to result from magnetic reconnection in the outer convective envelopes of these stars [13]. Magnetic reconnections are generally associated with sunspot/starspot pairs or groups, so the presence of flaring in stars is a reliable indicator of the presence of starspots (as well as solar coronae). A recent Kepler catalog study detected flaring stars in 4041 stars or 1.9% of the Kepler database, with the majority of these being of cool spectral type (F, G, K and M) [14]. Because hot stars (O, B, A) lack outer convective envelopes, they are not expected to support sunspots, magnetic reconnections, or stellar flares.

An early analysis of initial Kepler data revealed the surprising discovery of evidence for flaring in Kepler-discovered A stars [15,16], and additional evidence in Kepler data for the existence of exoplanets in orbit around rapidly rotating, spotted A stars [17]. More recent analysis of long cadence Kepler data has confirmed the existence of flaring in numerous A stars [18], thereby indicating the possibility of the existence of magnetic reconnection processes, solar coronae, and starspots in hot stars. The ability of SII observations to image in the U/V band allows high contrast imaging of cooler surface features, such as starspots, in A stars (and hotter). This unique high resolution imaging ability provides an important tool for the exploration of this poorly understood, unanticipated magnetic phenomena in hot stars.

**Rapidly Rotating Stars** Rapidly rotating stars are normally hot, of spectral types O, B, and A, making them well suited for imaging by SII. Some stars are rotating fast enough to reduce effective gravity at their equator. Fast rotation enables mass loss through the formation of circumstellar disks and hot polar winds, such as those in Be stars [19]. Rapid rotation also causes the star itself to become oblate, and induces gravity darkening. Spectral-line broadening reveals many early-type stars as rapid rotators. Although a combination of rapid and differential rotation is expected to occur in certain stars, it has yet to be observed. Initial imaging of rapidly rotating stars using amplitude interferometry has already revealed puzzling behavior; the rapid rotator Altair "shows stronger darkening along the equator, inconsistent with any von Zeipel-like gravity darkening prescription assuming uniform rotation" [20]. Sampling a large number of interferometer baselines enables the formation of model-independent images, which consequently also generates the capability to confirm or evolve our current theories of stellar evolution [21].

**Circumstellar disks** Rapid rotation lowers the effective gravity near the stellar equator which enables centrifugally driven mass loss and the development of circumstellar structures. Be-stars, containing Hα and other emission lines, define a class of rapid rotators with dense equatorial gas disks. Observations indicate the coexistence of the disk with a variable stellar wind at higher latitudes, and the disks may evolve, develop and disappear over timescales of months or years. Some Be-stars show outbursts, where the triggering mechanism may be coupled to non-radial pulsations [22]. A related group is B[e] stars (e.g., HD 62623 = l Pup, $m_V$ = 4.0), where emission is observed in forbidden atomic lines from [Fe II] and other species. As many of these circumstellar spectral lines are confined to longer wavelengths ( > 600 nm), U/V band imaging using kilometer baseline SII may provide an opportunity to separate the hot stellar components from the circumstellar disk environment.

**Winds from hot stars** The hottest and most massive stars (O-, B-, and Wolf-Rayet types) have strong and fast stellar winds that are radiatively driven by the strong photospheric flux being absorbed or scattered in spectral lines formed in the denser wind regions. Not surprisingly, their complex time variability is not well understood. Stellar winds can create co-rotating structures in the circumstellar flow similar to structures observed in the solar wind [23]. These structures are potentially responsible for discrete absorption components observed in ultraviolet P Cygni-type line spectra.

**Wolf-Rayet Stars and their environments** $\gamma^2$ Velorum is an ideal object for studies of circumstellar interactions, as it is the closest and brightest Wolf-Rayet star[24], and it resides in a binary with a hot O-type star [25]. The dense Wolf-Rayet wind collides with the less dense but faster O-star wind, generating shocked collision zones, wind-blown cavities and eclipses of spectral lines emitted from a probably clumpy wind. Bright V band emission lines motivate SII imaging studies in different passbands; the circumstellar emission region (seen in the C III-IV feature around λ= 465 nm) was found by NSII to be extended compared to the continuum flux from the stellar photosphere, filling much of the Roche lobe between the two binary components. The colliding wind binary (CWB) WR 140 ($m_V$ = 6.9, with bright emission lines) is another prime target for SII observations, where the hydrodynamic bow shock has been followed with milliarcsecond resolution by VLBA, revealing rotation as the orbit progresses during its 7.9 yr period. It is worth notiing that although WR 140 has not been detected at GeV energies [26], it remains the prime candidate for the study emission of VHE gamma-ray emission in CWB. Consequently, SII imaging of WR140 can proceed symbiotically with CTA observations of the VHE emission, providing an opportunity for truly simultaneous U/V band imaging in coincidence with the VHE gamma-ray observations.

**Blue Supergiants and Related Stars** Luminous blue variables occupy positions in the Hertzsprung-Russell diagram adjacent to those of Wolf-Rayet stars, and are suitable targets for SII imaging. Luminous blue variables possess powerful stellar winds and are often believed to be the progenitors of nitrogen-rich WR-stars. Rigel (β Ori; B8 Iab), the closest blue supergiant (240 pc, θ ~2.7 mas), is a very dynamic object that exhibits variable absorption/emission lines and oscillations on many different timescales. The properties of Rigel are known to resemble those of the progenitor to supernova SN1987A. The remarkable luminous blue variable η Carinae is the most luminous star known in the Galaxy. It is an extremely unstable and complex binary object that has undergone giant eruptions with huge mass ejections during past centuries. The mechanisms behind these eruptions are not understood but, like Rigel, η Car may well be on the verge of exploding as a core-collapse supernova. Interferometric studies can reveal asymmetries in the stellar winds [27] with enhanced mass loss along the rotation axis [28] (i.e., from the poles), resulting from the enhanced temperature at the poles that develops in rapidly rotating stars. Features at the angular size of the component stars are only resolvable with km scale baselines.

**Binary Systems** Most of the stars in the galaxy are binary or multiple star systems. Many of these systems have separations that are estimated only through spectroscopic observations. The procedures already described for measuring uniform disks and limb darkening assume we are dealing with a single star. In multiple star systems the correlation function is a more complicated function, where the correlation is a superposition of the components that will vary with time based on the angle the components make to the baseline. The orbital parameters can be determined by precisely measuring the change in interferometric visibility over time, given sufficient Fourier plane coverage (e.g. β Lyrae [29]). Extended SII observation campaigns can provide sufficient image resolution to reconstruct images of short period (days-months) binary systems over a full orbit, thereby exploring the roles of mass exchange and Roche lobe overflow in the evolution of binary systems, accretion discs, and High Mass X-ray Binaries [30].

**References**

[1] G.D. Nelson & S. Musman, "The Scale of Solar Granulation" *ApJ* 222:69, 1978.

[2] R.L. Bellot & D. Orozco Suárez, "Quiet Sun magnetic fields: an observational view" *LRSP* 16:1, 2019.

[3] R. Hanbury Brown, J. Davis, and L. R. Allen, "The Angular Diameters of 32 Stars." *MNRAS,* 167:121-136, April 1974.

[4] B.S. Acharya, et al., "Introducing the CTA Concept," *Astropart. Phys,* 43:3, 2013.

[5] The CTA Consortium, "Science with the Cherenkov Telescope Array," World Scientific, 2019.

[6] D. Dravins, "Intensity interferometry: optical imaging with kilometer baselines." *Optical and Infrared Interferometry and Imaging V*, volume 9907 of *Proceedings SPIE*, July 2016.

[7] P. D. Nunez, R. Holmes, D. Kieda, J. Rou, and S. LeBohec, "Imaging submilliarcsecond stellar features with intensity interferometry using air Cherenkov telescope arrays." *MNRAS,* 424:10061011, August 2012.

[8] G. Duvert, "VizieR Online Data Catalog: JMDC : JMMC Measured Stellar Diameters Catalogue." *VizieR Online Data Catalog*, 2345, November 2016.

[9] T. Boyajian, K. von Braun, G. A. Feiden, D. Huber, S. Basu, P. Demarque, D. A. Fischer, G. Schaefer, A. W. Mann, T. R. White, V. Maestro, J. Brewer, C. B. Lamell, F. Spada, M. Lopez-Morales, M. Ireland, C. Farrington, G. T. van Belle, S. R. Kane, J. Jones, T. A. ten Brummelaar, D. R. Ciardi, H. A. McAlister, S. Ridgway, P. J. Goldfinger, N. H. Turner, and L. Sturmann, "Stellar diameters and temperatures - VI. High angular resolution measurements of the transiting exoplanet host stars HD 189733 and HD 209458 and implications for models of cool dwarfs." *MNRAS,* 447:846–857, February 2015.

[10] S. Csizmadia, T. Pasternacki, C. Dreyer, J. Cabrera, A. Erikson, and H. Rauer, "The effect of stellar limb darkening values on the accuracy of the planet radii derived from photometric transit observations." *A&A*, 549:A9, January 2013.

[11] H. R. Neilson, J. T. McNeil, R. Ignace, and J. B. Lester, "Limb Darkening and Planetary Transits: Testing Center-to-limb Intensity Variations and Limb-darkening Directly from Model Stellar Atmospheres." *ApJ*, 845:65, August 2017.

[12] N. Espinoza and A. Jordan, "Limb darkening and exoplanets: testing stellar model atmospheres and identifying biases in transit parameters." *MNRAS,* 450:1879–1899, June 2015.

[13] B. R. Pettersen, "A review of stellar flares and their characteristics." IAU, Colloquium on Solar and Stellar Flares, 104th, Stanford, CA, Aug. 15-19, 1988 Solar Physics (ISSN 0038-0938), 121:299-312 1989.

[14] J. R. A. Davenport, "The Kepler Catalog of Stellar Flares." *ApJ,* 829:23, Sept 2016.

[15] L. A. Balona, "Kepler observations of flaring in A–F type stars." *MNRAS,* 423:3420, July 2012.

[16] L. A. Balona, "Activity in A-type stars." *MNRAS*, 431:2240, March 2013.

[17] L. A. Balona, "Possible planets around A stars." *MNRAS,* 441:3542, May 2014.

[18] T. V. Doorsselaere, H. Shariati, and J. Debosscher, "Stellar Flares Observed in Long-cadence Data from the Kepler Mission." *ApJ Supp.,* 232:26, October 2017.

[19] P. Kervella and A. Domiciano de Souza, "The polar wind of the fast rotating Be star Achernar. VINCI/VLTI interferometric observations of an elongated polar envelope."

*A&A* 453:1059-1066, July 2006.

[20] H. R. Neilson, J. T. McNeil, R. Ignace, and J. B. Lester, "Limb Darkening and Planetary Transits: Testing Center-to-limb Intensity Variations and Limb-darkening Directly from Model Stellar Atmospheres." *ApJ,* 845:65, August 2017.

[21] J. D. Monnier, M. Zhao, E. Pedretti, N. Thureau, M. Ireland, P. Muirhead, J.-P. Berger, R. Millan-Gabet, G. Van Belle, T. ten Brummelaar, H. McAlister, S. Ridgway, N. Turner, L. Sturmann, J. Sturmann, and D. Berger. "Imaging the Surface of Altair." *Science,* 317:342, July 2007.

[22] J.M. Porter and T. Rivinius. "Classical Be Stars.", *PASP* 115:1153-1170, October 2003.
Winds from hot stars:

[23] A. ud-Doula and S. Owocki. "Dynamical Simulations of Magnetically Channeled Line-driven Stellar Winds. I. Isothermal, Nonrotating, Radially Driven Flow." *ApJ* 576:413-428 September 2002.

[24] F. Millour et al. , "Direct constraint on the distance of $\gamma^2$ Velorum from AMBER/VLTI observations." *A&A* 464:107-118, March 2007.

[25] North et al. , "$\gamma^2$ Velorum: orbital solution and fundamental parameter determination with SUSI." *MNRAS* 377:415-424, May 2007.

[26] M. Werner, O. Reimer, A. Reimer and K. Egberts. "*Fermi*-LAT upper limits on gamma-ray emission from colliding wind binaries." *A&A* 555:A102, July 2013.

[27] van Boekel et al., "Direct measurement of the size and shape of the present-day stellar wind of eta Carinae." *A&A* 410:L37-L40, October 2003.

[28] Weigelt et al. "VLTI-AMBER velocity-resolved aperture-synthesis imaging of *η* Carinae with a spectral resolution of 12 000: Studies of the primary star wind and innermost wind-wind collision zone." *A&A* 594:A106, October 2016.

[29] Zhao et al. "First Resolved Images of the Eclipsing and Interacting Binary β Lyrae." *ApJ* 684:L95, September 2008.

[30] I. Negueruela, "Stellar Wind Accretion in High-Mass X-Ray Binaries.", in *"High Energy Phenomena in Massive Stars"* ASP Conference Series, 422:57, May 2010.

**Endorsements** (names and institutions):

Ivan Agudo, *IAA-CSIC, Spain*

Giovanni Bonanno, *INAF, Italy*

Kai Brügge, *TU Dortmund, Germany*

Sylvain Chaty, *University Paris Diderot CEA Saclay*

Paolo Coppi, *Yale University, New Haven, USA*

Filippo D'Ammando, *INAF-IRA Bologna, Italy*

Sebastian Diebold, Institute for Astronomy and Astrophysics, Eberhard Karls University Tübingen, Germany

Emma de Ona Wilhelmi, *DESY-Zeuthen & IEEC-CSIC, Germany*

Qi Feng, *Barnard College, New York, USA*

Tim Greenshaw, *University of Liverpool, UK*

Bohdan Hnatyk, *Astronomical Observatory of Taras Shevchenko National University of Kyiv, Ukraine*

Jon Lapington, *University of Leicester, UK*

Alexandre Marcowith, *Laboratoire Univers et Particules de Montpellier, France*

Giampero Naletto, *University of Padova, Italy*

Marek Nikolajuk, *University of Bialystok, Poland*

Michal Ostrowski, *Jagiellonian University, Poland*

Marco Roncadelli, *INFN-Pavia, Italy*

Guiseppe Romeo, INAF, *Osservatorio Astrofisico di Catania, Italy*

Olga Sergijenko, *Astronomical Observatory of Taras Shevchenko National University of Kyiv, Ukraine*

Justin Vendenbroucke, *University of Wisconsin-Madison, USA*

Jamie Williams, *University of Leicester, UK*